\documentclass[dvips]{article}
\usepackage{icrctc07}

\title{Magnetic turbulence production by streaming cosmic rays upstream of SNR shocks}
\shorttitle{Magnetic turbulence production at SNR shocks}
\authors{J. Niemiec$^{1,2}$, M. Pohl$^{1}$.}
\shortauthors{Niemiec and et al}
\afiliations{$^1$Department of Physics and Astronomy, Iowa State University, Ames, 
IA 50011, USA\\ $^2$Instytut Fizyki J\c{a}drowej PAN, ul. Radzikowskiego 152,
 31-342 Krak\'{o}w, Poland}
\email{niemiec@iastate.edu}

\abstract{We present preliminary results of Particle-In-Cell simulations of 
magnetic turbulence production by isotropic cosmic-ray ions streaming 
upstream of supernova remnant shocks. The studies aim at testing the MHD 
predictions by Bell \cite{bell04,bell05} of a strong amplification of short-wavelength 
nonresonant wave modes and at studying the subsequent evolution of the 
magnetic turbulence and its backreaction on cosmic ray trajectories. The detailed 
knowledge of the upstream turbulence properties is crucial to ascertain all 
aspects of the shock acceleration process - the transport properties of cosmic 
rays, the shock structure, thermal particle injection and heating processes. 
An amplification of magnetic field would also facilitate the acceleration of 
particles beyond the "knee" in the cosmic-ray spectrum. Our kinetic approach 
is particularly suited to address the backreaction on the cosmic rays, and it 
allows us to test Bell's prediction of the eventual formation of extended 
filamentary structure in the cosmic-ray distribution and also to arrive at 
a reliable estimate of the total saturation magnetic-field level. The 
parameters chosen for the simulations are favorable for the rapid excitation 
of purely growing modes. We confirm the generation of the turbulent magnetic
field due to the drift of cosmic-ray ions in the upstream plasma, but show that
the growth rate of the field perturbations is much 
slower than estimated using the MHD approach and the amplitude of the turbulence
saturates at about $\delta B/B\sim 1$. The magnetic field also remains below
equipartition with the upstream plasma.}

\begin{document}
\maketitle

\section{Introduction}
Supernova remnant shock waves are prime candidates for the acceleration sites of 
Galactic cosmic rays. However, no firm evidence has been found so far for hadron 
production in SNRs, which challenges our understanding of these sources. Particle 
confinement near the shock is supported by self-generated magnetic turbulence and 
thus the detailed knowledge of the properties of the turbulence is crucial to 
ascertain all aspects 
of the acceleration process –-- transport properties of cosmic rays, the shock 
structure, thermal particle injection and heating processes. Of particular 
interest is the question of how efficiently and with what properties 
electromagnetic turbulence is produced by energetic particles at some distance 
upstream of the forward shocks. As the flow convects that turbulence toward and 
through the shock, the turbulent magnetic field structure at and behind the shock 
is shaped by the plasma interactions ahead of it.

We investigate the properties of magnetic turbulence upstream of the shocks of 
young SNRs. Cosmic rays accelerated at these shocks 
stream relative to the upstream plasma with the shock velocity, with the highest 
energy particles propagating the farthest from the shock and forming a nearly 
monoenergetic quasi-isotropic population at some distance upstream. Such a system has 
been recently studied by Bell \cite{bell04} with a quasilinear MHD approach. Bell noted 
that the current carried by drifting cosmic rays excites non-resonant, nearly 
purely growing modes more efficiently than resonant Alfv\'en waves, which he 
verified with MHD simulations. Later studies suggested that regions of low 
density and  low magnetic  field  strength
would be formed, thus resembling a filamentation instability \cite{bell05}. 
Here we use Particle-In-Cell simulations to verify 
the existence of the non-resonant modes in the presence of kinetic 
particle effects and explore its properties such as nonlinear evolution, 
saturation, and particle heating.

\begin{figure*}[t]
\mbox{\includegraphics[width=7.1cm]{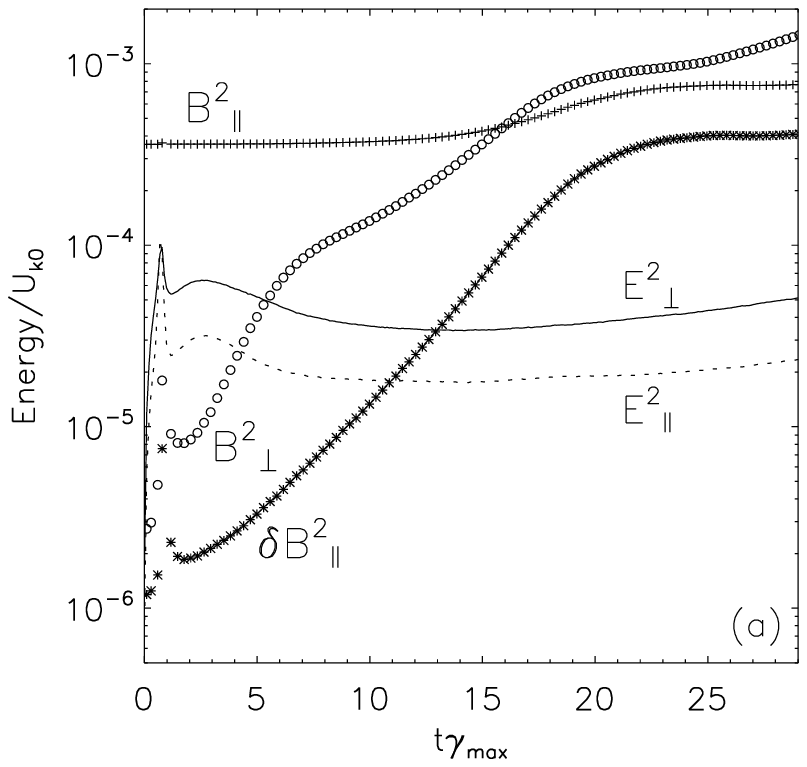} \hspace*{0.55cm}
      \includegraphics[width=7.1cm]{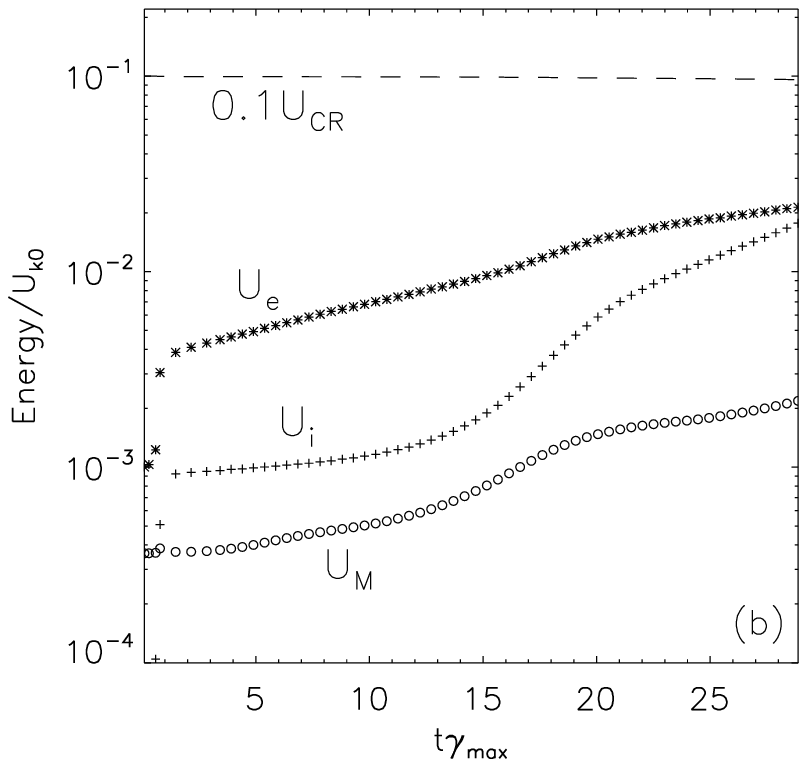}}
\caption{Temporal evolution of the total energy density in fields (a) and particles 
 ($U_{CR}, U_i, U_e$) and the magnetic field $U_M$ (b) normalized to the 
 initial total kinetic energy in the system $U_{k0}$.
 Time is in units of the growth rate $\gamma_{max}$ for the most unstable mode 
 with wavelength $\lambda_{max}$. 
 $B^2_{\perp}$ and  $E^2_{\perp}$ indicate the magnetic and electric field energy
 densities in the
 components perpendicular to the cosmic-ray ion drift direction. 
 The total kinetic energy in cosmic-ray ions $U_{CR}$ is 
 rescaled by a factor of 10.}
\end{figure*}
\section{Particle-in-cell experiment and results}
We use a modified version of the relativistic 3D code TRISTAN.  
In the simulations, an isotropic population of relativistic, monoenergetic cosmic-ray
ions with Lorentz factor $\gamma_{CR}=2$ and number density $N_{CR}$ drifts with 
$v_{sh}=0.3c$
relative to the upstream electron-ion plasma and along the homogeneous magnetic 
field $B_{\parallel 0}$, carrying a current density $j_{CR}=eN_{CR}v_{sh}$. 
The ions of the upstream medium have a thermal distribution
with number density $N_i$, in thermal equilibrium with the electrons. The electron 
population with density $N_e=N_i+N_{CR}$ 
contains the excess electrons to provide charge-neutrality and also drifts with 
$v_d=v_{sh}N_{CR}/N_e$ with
respect to the upstream ions providing a return current $j_{ret}=-eN_ev_d$
to balance the current in 
cosmic rays. The density ratio assumed in this study is $N_{CR}/N_i$=1/3.
The simulations have been performed using a 992$\times$304$\times$304 computational 
grid with periodic boundary conditions and
with a total of 1.5 billion particles. Cosmic-ray ions drift in the $-x$-direction.
The electron skindepth $\lambda_{se}=c/\omega_{pe}=7\Delta$, where 
$\omega_{pe}$ is the electron plasma frequency
and $\Delta$ is the grid cell size, and the ion-electron mass ratio assumed $m_i/m_e=10$.
The above parameter combination was chosen to resemble physical conditions in 
young SNRs and, according to quasilinear calculations \cite{bell04}, is favorable 
for the rapid excitation of purely growing, short-wavelength (as compared with 
cosmic-ray ion gyroradii $r_{CRg}$) wave modes. In 
the unstable wavevector regime \mbox{$1 < k_\parallel r_{CRg} < \zeta v_{sh}^2/v_{A}^2$}, the 
dispersion relation for the purely growing mode
\begin{equation} 
\omega^2 - v_{A}^2 k^2_\parallel + \zeta v_{sh}^2 k_\parallel/r_{CRg} = 0 
\end{equation}
gives the maximum growth rate $\gamma_{max}$ at wavenumber  
$k_{_\parallel max}$ = $\zeta v_{sh}^2/2v_{A}^2r_{CRg}$, where $k_\parallel$ is wavevector
parallel to $B_{\parallel 0}$, 
$v_{A}$ is the Alfv\'en velocity,
and $\zeta=N_{CR}c\gamma_{CR}/N_iv_{sh}$. Our parameters give 
$\zeta v_{sh}^2/v_{A}^2\simeq 826$ 
and the most unstable mode with wavelength $\lambda_{max}=50\Delta$ is in the 
middle of the unstable wavevector range.  

\begin{figure*}[th]
\begin{center}
\includegraphics[width=14.8cm]{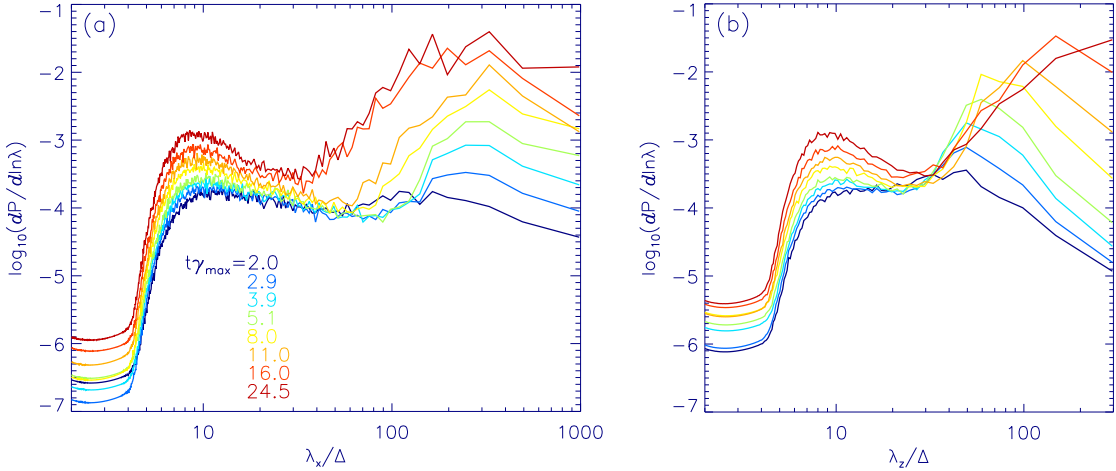} 
\end{center}
\caption{Power spectra of the magnetic field component $B_y$ in the direction
parallel (a) and perpendicular (b) to the cosmic-ray ions drift direction at 
different times as given in units of the growth rate $\gamma_{max}$. 
The spectra were calculated in the 2D slice in the $xz$-plane at $y/\Delta=200$.
}
\end{figure*}

Figure 1 shows the temporal evolution of the magnetic and electric field and 
particle kinetic energy
densities. An initial (up to $t\sim\gamma_{max}^{-1}$) fast growth of the turbulent 
fields is caused by a Buneman-type beam instability
between slowly drifting electrons and the ambient ions, which leads to 
plasma heating until the electron thermal velocities become comparable
to the electron drift speed $v_d$. In real SNRs the density ratio is likely
much larger than the sound Mach number, and therefore the electron drift speed
is smaller than the thermal velocity. The initial instability is thus solely
a consequence of the initial conditions in our simulations, in particular the
initial temperature of the background medium.

The wave modes due to the instability caused by the cosmic-ray ions streaming
in the upstream plasma start to emerge at $t\sim 2\gamma_{max}^{-1}$ and the
magnetic field turbulence is excited mainly in the components transverse 
to the cosmic-ray drift direction. The dominant wave mode is oblique at an
angle of about 75$^o$ to the $x$-direction (Fig. 2), and has the wavelength 
component $\lambda_\perp \approx 50\Delta$, which is numerically similar to $\lambda_{max}$.
The character of the mode is thus different than that predicted from the linear analysis 
by Bell \cite{bell04,bell05}, which assumed that the most rapid growth occurs for
wavenumbers parallel to $B_{\parallel 0}$ (Eq. 1). As seen in Fig. 2, the dominant scale
in $\lambda_\perp$ grows as the turbulence develops, but the parallel scales 
($\lambda_\parallel=\lambda_x$) remain roughly constant, so after about 
$20\gamma_{max}^{-1}$ the structures have similar size parallel and perpendicular
to the drift. Comparison with Fig. 1 reveals that at those late times 
$\delta B_\perp$ is of the order of $B_{\parallel 0}$, so the 
system becomes magnetically nearly isotropic. The drift of the cosmic-ray particles
is still kinetically dominant, which suggests an important role of the 
homogeneous magnetic field component for the evolution of the magnetic turbulence.

The spatial structures of the turbulent magnetic field and accompanying ambient
plasma distribution are shown in Fig. 3 for $t\approx 10\gamma_{max}^{-1}$. 
The perpendicular field is concentrated around regions of low background 
plasma density. The cosmic-ray distribution is homogeneous, and therefore the 
magnetic field is associated with a strong net cosmic-ray current flowing inside 
ambient-plasma voids, and the field lines circle around the center of the
cavities according to Ampere's law. The formation of the cavities in the plasma 
is due to the $\vec{j}_{ret}\times\vec{B}$ force, which accelerates the upstream
plasma away from the center of the voids, and also causes the cavities to expand.
This is visible as the evolution towards larger scales shown in Fig.~2b. 
The growth of the cavities and the subsequent merging of the adjacent plasma voids
lead to a compression of the plasma between cavities and also to an amplification
of the magnetic field. Because the magnetic field has a preferred orientation around
each cavity, the magnetic field lines cancel each other in the space between the
voids (Fig. 3).   

\begin{figure*}[t]
\mbox{\includegraphics[width=7.1cm]{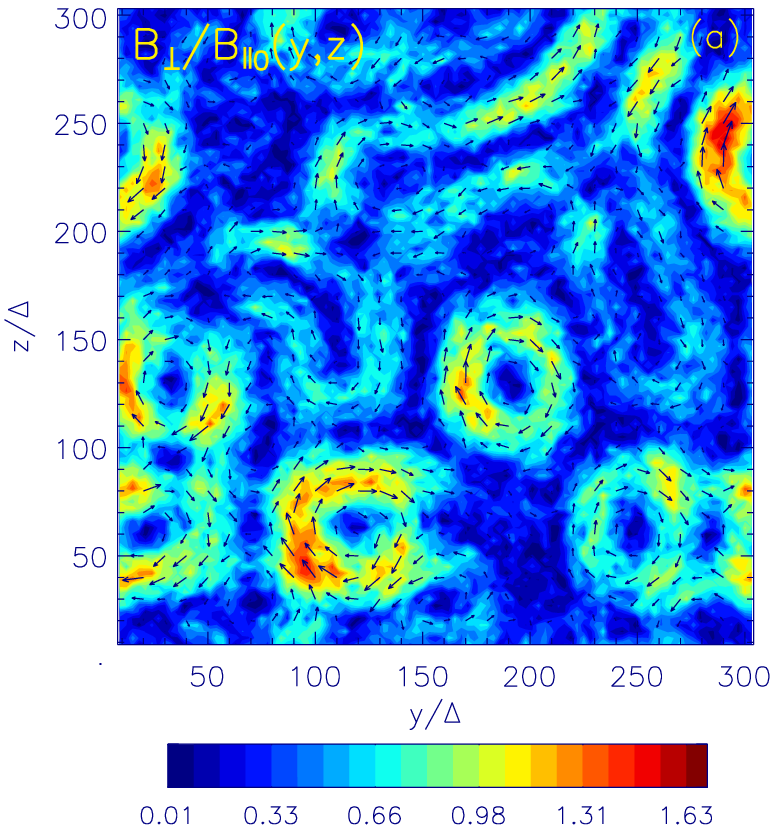} \hspace*{0.55cm}
      \includegraphics[width=7.1cm]{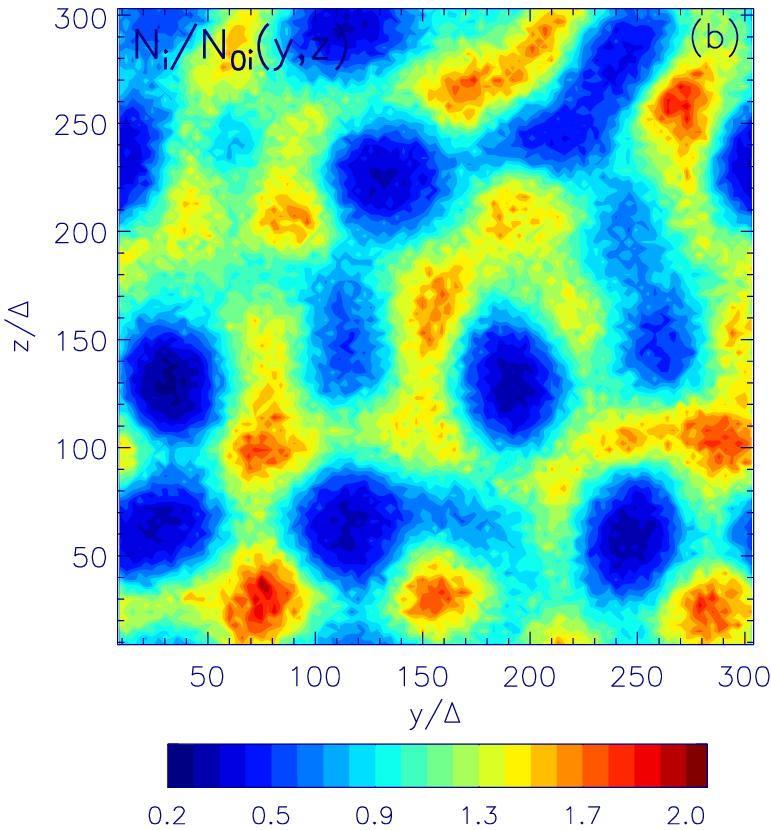}}
\caption{The magnitude and direction of the perpendicular magnetic field component 
 $B_{\perp}=(B^2_y+B^2_z)^{1/2}$ (a) and the ambient ion density $N_i$ (b) in the
 plane perpendicular to the cosmic-ray ion drift direction at $x/\Delta=500$
 and $t\approx 10\gamma_{max}^{-1}$. $B_{\perp}$ is normalized to the amplitude
 of the homogeneous field $B_{\parallel 0}$, and $N_i$ to the initial
 ambient ions density. The electron distribution follows that of ambient ions.}\label{Fig:CL}
\end{figure*}

Compared with Bell's MHD studies, the growth of the magnetic 
perturbations in our kinetic modeling is much slower. The initial 
growth rate
($t\sim (2-6) \gamma_{max}^{-1}$) 
of the perpendicular field turbulence is only 1/5 of $\gamma_{max}$, and becomes
smaller during the later evolution. 
Moreover, the amplitude of the turbulent component never 
considerably exceeds the amplitude of the regular field and the field growth
saturates when $\delta B_\perp/B_{\parallel 0}\approx 1$. 
Note, that although for $t > 25 \gamma_{max}^{-1}$ the field structures become
larger than the size of the simulation box, the magnetic field saturation level
is not influenced by the boundary conditions, which we have confirmed using
larger-box 2D simulations, that well reproduce the results of the 3D experiment. 
The field growth is
also accompanied by heating  of the ambient plasma, and the field remains below 
equipartition with the upstream medium.  
The cosmic-ray density 
distribution remains roughly unchanged till the end of the simulation. 
\section{Summary and conclusions}
The process of the production of magnetic field turbulence by cosmic-ray ions
drifting upstream of SNR shocks has been studied here using PIC simulations.
Turbulent field is indeed generated in this process, but
the growth of magnetic turbulence is much 
slower than estimated using the MHD approach, and the amplitude of the field
perturbations saturates at approximately the amplitude of the homogeneous 
upstream field. The energy density in the turbulent field is also
always much smaller than the plasma kinetic energy density. This suggests that
the efficiency of magnetic field generation in this mechanism may not be sufficient
to account for the high magnetic fields observed in some young SNRs and also
leaves open the question of the ability of diffusive particle acceleration 
at SNR shocks to produce particles with energies beyond 
the "knee" in the cosmic-ray spectrum.

{\small \noindent
The simulations have been performed on
Columbia at NAS and Tungsten at NCSA which
is supported by the NSF. The work was supported by MNiSW during 2005-2008 as 
research project 1 P03D 003 29.}
\bibliography{icrc1047}

\begin{thebibliography}{1}

\bibitem{bell04}
A.~R. {Bell}.
\newblock {Turbulent amplification of magnetic field and diffusive shock
  acceleration of cosmic rays}.
\newblock {\em MNRAS}, 353:550--558, 2004.

\bibitem{bell05}
A.~R. {Bell}.
\newblock {The interaction of cosmic rays and magnetized plasma}.
\newblock {\em MNRAS}, 358:181--187, 2005.

\end{thebibliography}
\bibliographystyle{plain}

\end{document}